\documentclass[twocolumn,prb,floatfix,amsmath,amssymb,authordate1-4]{revtex4}

\usepackage[dvips]{graphicx}
\usepackage{dcolumn}
\usepackage{bm}
\usepackage{epsfig}

\begin{document}


\title{Molecular Dynamics Study of Charged Dendrimers in Salt-Free Solution:\\
Effect of Counterions
}

\author{Andrey A. Gurtovenko}
\affiliation{Laboratory of Physics and Helsinki Institute of Physics, 
Helsinki University of Technology, P.O.Box 1100, FIN-02015 HUT, Finland~ and\\
Institute of Macromolecular Compounds, Russian Academy of Sciences,
Bolshoi Prospect 31, V.O., St.Petersburg, 199004 Russia
}

\author{Sergey V. Lyulin}
\affiliation{Institute of Macromolecular Compounds, Russian Academy of Sciences,
Bolshoi Prospect 31, V.O., St.Petersburg, 199004 Russia
}

\author{Mikko Karttunen}
\affiliation{Biophysics and Statistical Mechanics Group,
Laboratory of Computational Engineering, Helsinki University
of Technology, P.\,O. Box 9203, FI--02015 HUT, Finland}

\author{Ilpo Vattulainen}
\affiliation{Laboratory of Physics and Helsinki Institute of Physics, 
Helsinki University of Technology, P.\,O. Box 1100, FI--02015 HUT, Finland, 
and Memphys--Center for Biomembrane Physics, Physics Department, 
University of Southern Denmark, Campusvej 55, DK--5230 Odense M, Denmark}

\date{\today}

\begin{abstract}
Polyamidoamine (PAMAM) dendrimers, being protonated under 
physiological conditions, represent a promising class of 
nonviral, nano-sized vectors for drug and gene delivery. 
We performed extensive molecular dynamics simulations of 
a generic model dendrimer in a salt-free solution with 
dendrimer's terminal beads
positively charged. Solvent molecules as well as counterions 
were {\it explicitly} included as interacting beads. We find 
that the size of the charged dendrimer depends 
{\it non-monotonically} on the strength of electrostatic 
interactions demonstrating a maximum when the Bjerrum length 
equals the diameter of a bead. Many other structural and dynamic 
characteristics of charged dendrimers are also found to follow 
this pattern. We address such a behavior to the interplay between 
repulsive interactions of the charged terminal beads and their 
attractive interactions with oppositely charged counterions. 
The former favors swelling at small Bjerrum lengths and the 
latter promotes counterion condensation. Thus, counterions 
can have a dramatic effect on the structure and dynamics of 
charged dendrimers and, under certain conditions, cannot be 
treated implicitly.
\end{abstract}

\maketitle

\section{\label{Intro}Introduction\protect }

Dendrimers, characterized by regular branching with radial 
symmetry, represent a unique class of polymers possessing 
a large number of functional terminal groups in their outermost 
dendritic shell.\cite{Frechet94} These ``dream molecules''\cite{Service95} 
have recently attracted a lot of attention 
due to potential applications in technology and medicine, see, 
for example, Refs.~\onlinecite{Jaaskelainen:00js,Boas:03kr,Ozkan:04cn} 
and references therein. Polyamidoamine (PAMAM) dendrimers, 
in particular, are commonly used molecules in this field. 
They are protonated under physiological conditions,\cite{Duij98,Jaaskelainen:00js} 
thus their high surface charge 
coupled to their ability to adapt conformation in response to 
changes in their surrounding environment make PAMAM dendrimers 
very attractive candidates for biomedical applications such as 
drug and gene 
delivery.\cite{Kukow96,Biel96,Liu99,Jaaskelainen:00js,Essfand01,Clon02,Sonawane:03ev,Gillies05}
Thus far, most theoretical and computational investigations have focused on 
neutral dendrimers.\cite{Ball04,Gurt05a}

Since the pioneering paper of Welch and 
Muthukumar,\cite{Welch98} there have been only a few 
computational studies of charged model dendrimers.\cite{Welch98,LyulinS04,LyulinS04a,Terao04} 
While these studies have provided a great deal of insight 
into the properties of dendrimers, it is obvious that 
their scope has been limited due to a major computational 
cost associated with modeling of complex dendrimers in 
a hydrodynamic solvent. Hence, the previous 
studies have treated the groups of a charged dendrimer 
using the linearized Poisson--Boltzmann (Debye--H{\"u}ckel) 
theory, and the solvent as a continuous medium.
Despite these limitations, the above simulations demonstrated 
that the strength of electrostatic interactions 
may have a strong effect on the dendrimer's size as well as 
on other structural and dynamic properties. In particular, it 
was shown that the size of any charged dendrimer increases 
for an increasing effective screening length which corresponds 
to a decreasing salt concentration.\cite{Welch98,LyulinS04,LyulinS04a,Terao04}

As for more accurate treatments, Lee et al.\cite{Lee02} 
simulated PAMAM dendrimers at an atomic resolution. They 
found that a PAMAM dendrimer swells significantly when pH 
decreases from a high pH value (no amines are protonated) 
to neutral (all the primary amines are protonated), and to 
low pH (all the primary amines and tertiary amines are 
protonated). However, they treated the counterions as well 
as the solvent implicitly (except for dendrimers of second 
generation which were solvated in explicit water).\cite{Lee02} 
A more realistic atomic-scale simulation of PAMAM dendrimers 
was performed very recently by Goddard et al.,\cite{Maiti05,Lin:05pa} 
who accounted for water molecules and counterions in an explicit 
manner. They showed that the presence of solvent leads to 
a swelling of the dendrimer, while the effect of pH was found 
to be similar to that reported in earlier studies.\cite{Lee02} 
As a part of counterions were found to condense onto the 
dendrimer,\cite{Maiti05} it was stated that the presence 
of counterions increases the swelling of a dendrimer. The 
effect of counterions on dendrimer charges and related 
structural aspects were not clarified, though, since for the 
counterion concentrations they used, the amine charges were 
strongly screened: the corresponding Debye length was about 
3\,--\,4\,\AA\, and hence close to the minimal value considered 
by Welch and Muthukumar.\cite{Welch98} This implies that 
the effect of smaller counterion concentrations and, hence, 
the overall role of counterions has remained unresolved.

In general, there is reason to emphasize that the applicability 
of the Debye--H{\"u}ckel approximation for dendrimers should 
not be taken for granted. It is valid in the dilute limit--here 
it means the situation where the separation between charges 
exceeds the Bjerrum length. However, that condition breaks down 
easily in dense structures such as dendrimers, which implies that 
in most of the cases counterions must be taken {\it explicitly} 
into account. Computational studies of flexible linear 
polyelectrolytes have demonstrated\cite{Stevens95,Holm:01js,Liu02} 
that the presence of explicit counterions is indeed crucial: 
increasing the strength of electrostatic interactions 
can provoke counterion condensation and, therefore, lead to 
dramatic changes in the polymer's structure and dynamics. 
It is reasonable to expect similar counterion effects also 
for dendrimers. A recent computational study of model dendrimers 
suggests that this is indeed the case.\cite{Gal05}

In this study we aim to clarify the role of counterions. 
We have performed extensive molecular dynamics simulations 
using a generic cationic model dendrimer of the fourth generation. 
Counterions and solvent molecules are {\it explicitly} included. 
A systematic variation of the strength of electrostatic interactions 
have allowed us to examine the influence of counterions on the 
structural and dynamical properties of charged dendrimers.
It turns out that the size of charged dendrimers depends 
{\it non-monotonically} on the strength of electrostatic 
interactions in the system clearly indicating a pronounced 
effect of counterion condensation.

\section{\label{model}Model and Simulation Details\protect}

\begin{figure}[tb] 
\includegraphics[width=6.0cm,clip=true]{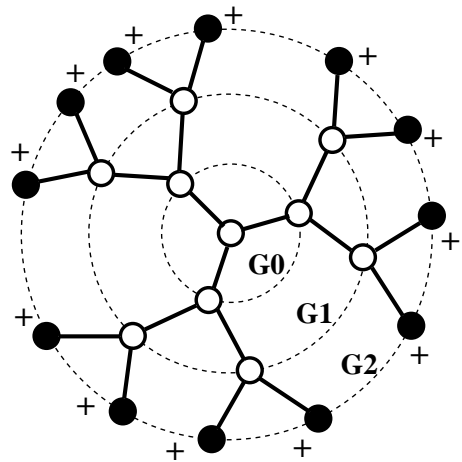}
\caption{\label{Dend}
 A sketch of the dendritic topology used in this study. 
 Beads of the last, or the terminal, generation shell 
 (the illustration is for a dendrimer of the second 
 generation) are positively charged (black). In the 
 simulations we used dendrimers of the fourth generation.}
\end{figure}

We performed MD simulations of dendrimers of the 4th generation 
using the freely jointed ``bead-and-spring'' model. Each bead of 
the last, or the terminal, generation was set to carry a positive 
unit charge. The dendrimer was solvated in a box of non-polar 
molecules (beads) and an appropriate number of counterions was 
added to preserve overall charge neutrality.

The bonds were modeled as harmonic springs with their potential 
energy given by $U = \frac{1}{2}k(l-l_{\rm 0})^2$, where 
$l_{\rm 0}$ is the reference bond length. The spring constant 
$k = 10^3\times k_{\rm B}T /l_0^2$ was chosen to be rather 
high as has been done in earlier studies.\cite{Karat01} All 
short-range non-electrostatic non-bonded interactions were 
described by the Lennard--Jones potential $U_{\rm LJ}=4\epsilon  
[(\sigma /r)^{12} - (\sigma /r)^{6}]$ with $\epsilon = 0.3\,k_{\rm B}T$.
The bead--bead interaction diameter $\sigma = 0.8\,l_0$ was 
taken to be the same for all beads in the system. 
The Lennard--Jones potential was cut off at $r_{\rm c}=2^{1/6}\sigma$ 
and shifted to zero.\cite{Stevens95}  The dendrimer was in 
a good solvent.

The long-range electrostatic interactions between the charged 
terminal beads (with a positive unit charge on each) and negatively 
charged counterions were handled using the particle-mesh Ewald 
(PME) method,\cite{Dar93,Ess95a} which has been shown to 
perform well in soft matter simulations.\cite{Pat03:pme,Pat04:pme} 
The strength of electrostatic interactions is characterized by 
the so-called Bjerrum length 
\begin{equation} 
\lambda _{\rm B} = \frac{e^2}
                   {4\pi \varepsilon_0 \varepsilon_{\rm s} k_{\rm B}T}~ , 
\end{equation}
where $\varepsilon_{\rm s}$ is the dielectric permittivity of 
the solvent. To examine the effects of counterions on the 
dendrimer's structure and dynamics, the Bjerrum length 
$\lambda_{\rm B}$ was systematically varied from 0 to $8.0\,l_0$. 
This change in $\lambda_{\rm B}$ may be interpreted as 
a variation of the dielectric properties of the solvent 
($\lambda_{\rm B}\sim 1/\varepsilon_s$). In all, we studied 
10 different systems with $\lambda_{\rm B}/l_0$ taken to be 
0.0; 0.2; 0.4; 0.6; 0.8; 1.0; 2.0; 3.2; 6.4; and 8.0.

The dendritic topology employed in our study is schematically 
depicted in Fig.~\ref{Dend}. We studied a trifunctional 
dendrimer of the fourth generation with single bonds 
between the branch points. The dendrimer had 94 beads of 
which 48 were terminal ones. To generate initial configurations, 
the procedure of Murat and Grest was used.\cite{Murat96} 
The dendrimer was then solvated in a cubic box with 4692 
solvent particles. That corresponds to a polymer fraction 
of about 0.02. The density was set to 1.688\,$l_0^{-3}$ 
(or, alternatively, to 0.864\,${\sigma}^{-3}$), similar to 
Refs.~\onlinecite{Dun93} and \onlinecite{Chang03}. The linear 
size of the simulation box size was set to 14.16\,$l_0$, and 
periodic boundary conditions were applied in all three 
dimensions. The ratio of the average dendrimer size to the 
simulation box size was about 0.2.

For simulating charged dendrimers, unit positive charges were 
assigned to all beads of the terminal G4 generation shell of 
the neutral dendrimer. This corresponds to physiological 
(neutral) pH conditions when the primary amines of a PAMAM 
dendrimer become protonated.\cite{Duij98} To keep the system 
electroneutral, randomly chosen 48 solvent beads were converted 
to counterions by assigning negative unit charges on them. 
This system was used as the starting configuration for MD 
simulations of charged dendrimers.

All systems with different Bjerrum lengths were first 
equilibrated for $10^6$ time steps in the NVT ensemble with 
the time step $\Delta t = 2.4\times 10^{-3}\,\tau_0$, where 
$\tau_0 = l_0 (m/\epsilon )^{1/2}$ is the characteristic 
time of the model ($m$ is the mass of a bead). Temperature 
was controlled using the Berendsen scheme,\cite{Ber84} while 
all the bond lengths were kept constant by using the SHAKE 
algorithm.\cite{Rick77} After equilibration, data were 
collected from production runs of $10^7$ time steps each in 
the NVE ensemble. The time step in the production runs was 
set to $1.2\times 10^{-3}\,\tau _0$ (constraints were not 
applied to dendrimer's bonds during the production). All the 
simulations were performed using the Gromacs package.\cite{Ber95,Lin01}  
Each simulation was run on a single 3.2~GHz Pentium-4 processor.

\begin{figure}[tb] 
\includegraphics[width=8cm,clip=true]{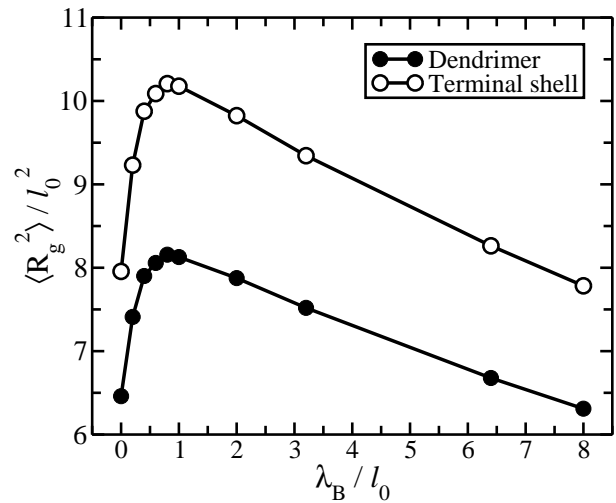}
\caption{\label{Rg} 
 Average mean-square radius of gyration 
 $\langle R_{\rm g}^2 \rangle$ of the whole dendrimer 
 (solid circles), and of the beads of the \textit{terminal} 
 dendrimer shell (open circles) as a function of the Bjerrum 
 length $\lambda_{\rm B}$.  Error bars are of the same
 size as the symbols. The error bars were estimated as the 
 standard errors of mean by splitting trajectories into 10 
 pieces of 10$^6$ time steps each.}
\end{figure}

\begin{figure}[tb] 
\includegraphics[width=6cm]{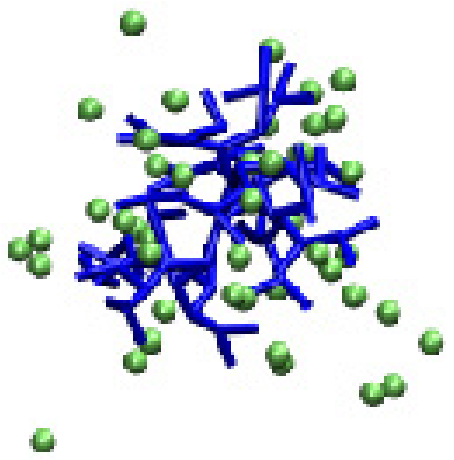}
\includegraphics[width=6cm]{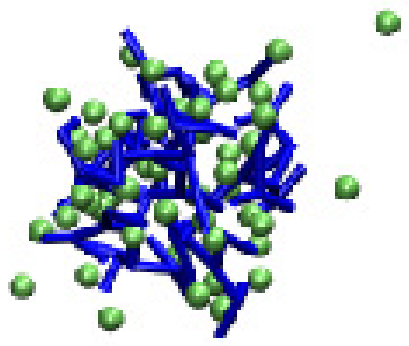}
\caption{\label{Snap}
 (Color online). Typical configurations of charged dendrimers at 
 $\lambda_{\rm B} = 0.8\,l_0$ (top) and $\lambda_{\rm B} = 6.4\,l_0$ 
 (bottom). Counterions are shown as spheres. Solvent is 
 omitted for clarity. 
}
\end{figure}

\begin{figure}[tb] 
\includegraphics[width=8cm,clip=true]{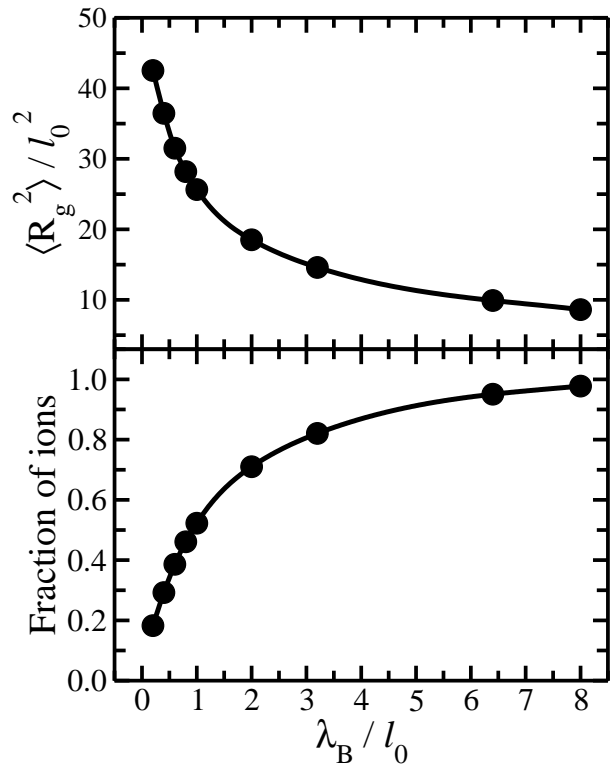}
\caption{\label{Ions} 
 {\it Top}: Average mean-square radius of gyration 
 $\langle R_{\rm g}^2 \rangle$  calculated for counterions
 with respect to the center of mass (CM) of the dendrimer.
 {\it Bottom}: Average fraction of counterions condensed onto 
 the charged dendrimer as a function of the Bjerrum length 
 $\lambda_{\rm B}$. See text for the criterion used to 
 identify condensed ions.}
\end{figure}

\begin{figure}[tb] 
\includegraphics[width=8cm,clip=true]{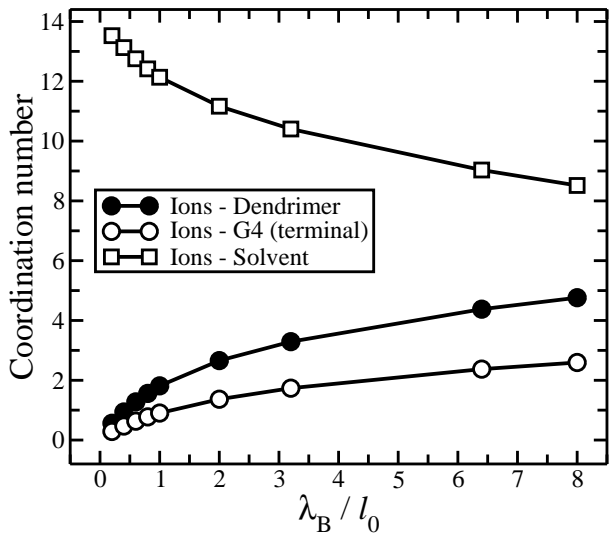}
\caption{\label{CN} 
 Average coordination numbers of ions with the dendrimer's beads
 (solid circles), with the charged beads of the G4 shell (open circles),
 and with solvent beads (open squares) as a function of the Bjerrum length 
 $\lambda _{\rm B}$.}
\end{figure}

To examine possible finite size effects, we performed additional 
simulation runs using simulations containing twice the number 
of solvent molecules. We examined the box size effects for 
a neutral dendrimer as well as for the charged dendrimers 
with $\lambda_{\rm B}=0.4\,l_0$ and $\lambda_{\rm B} = 6.4\,l_0$. 
For the case of neutral dendrimers we did not find any noticeable 
deviations when the larger box size was employed. For charged 
dendrimers, the average size of the counterion cloud around 
the dendrimer was found to naturally increase upon increasing 
the box size. This, however, did not produce any considerable 
changes in the structural or dynamic properties of dendrimers.

\section{\label{structure}Structural properties\protect}
\subsection{\label{equil}Equilibration\protect}

As in previous computational studies,\cite{Karat01,LyulinS04} 
initial equilibration was monitored through the time evolution 
of the mean-square radius of gyration
\begin{equation}
 R_{\rm g}^2  = \frac{1}{N}~ \sum _{i=1}^N (r_i - R_{\rm CM})^2, 
\end{equation}
where $N$ is the number of beads and $R_{\rm CM}$ is the 
position of the center of mass (CM) of the dendrimer. This 
quantity, being a measure of the overall size of the dendrimer, 
was found to remain stable during the entire production run.

The above, however, is not enough for a system containing 
unscreened charges. Is it further required that the ions 
have reached equilibrium. To monitor the equilibration of 
ions, we first calculated the radial distribution functions 
(RDFs) for ion--ion, ion--dendrimer bead, and ion--solvent 
molecule pairs (data not shown). The position of the first 
minimum defines the radius of the first coordination shell 
of a counterion. This procedure has been used in other 
related studies as well.\cite{Boeckman03,Gurt05,Gurt05salt} 
The radius was found to be $\sim$1.3\,$l_0$. The coordination 
numbers were then calculated by counting the total numbers 
of beads within the first coordination shell. The coordination 
numbers are found to become stable as a function of time during the
equilibration period (see Sect.~\ref{model}).

\subsection{\label{size}Dendrimer size and counterion condensation\protect}

Next, we describe the structural characteristics of charged 
dendrimers at different strengths of electrostatic interactions. 
In Fig.~\ref{Rg} we plot the average mean-square radius of 
gyration $\langle R_{\rm g}^2 \rangle$ of the dendrimer as 
a function of Bjerrum length $\lambda_{\rm B}$. Remarkably, 
the dependence of $\langle R_{\rm g}^2 \rangle$ on 
$\lambda_{\rm B}$ is found to be {\it non-monotonic}. 
Increasing the Bjerrum length from zero leads to a pronounced 
swelling of the dendrimer, the maximum being at 
$\lambda_{\rm B}\simeq 0.8\,l_0$ as shown in Fig.~\ref{Rg}. 
Further increase is found to have an opposite effect, i.e., 
the dendrimer shrinks. Such a behavior resembles the findings 
for linear\cite{Stevens95,Liu02} and dendritic\cite{Gal05} 
polyelectrolytes, and is associated with counterion condensation. 
Figure~\ref{Snap} shows snapshots from our simulations for  
the swollen and shrunk dendrimers.

\begin{figure*}[tb] 
\includegraphics[width=11cm,clip=true]{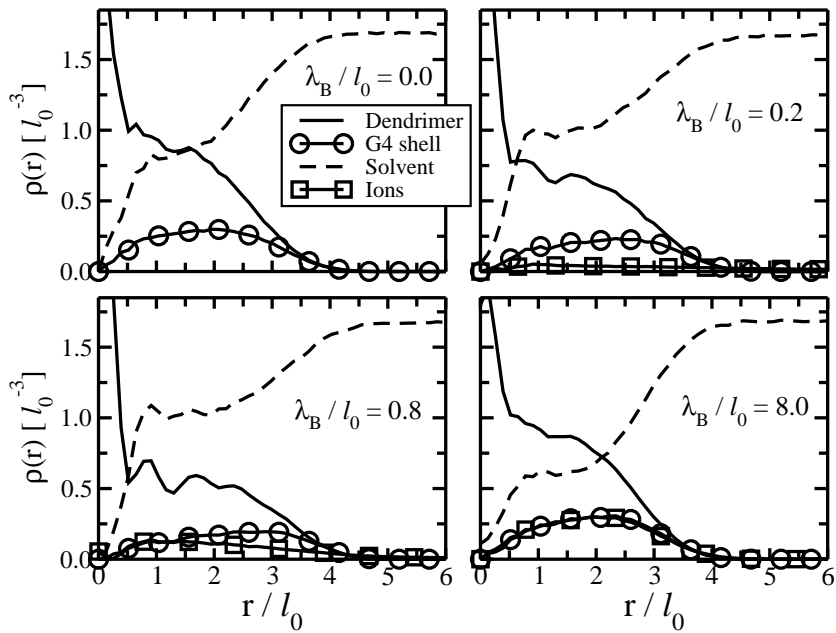}
\caption{\label{Density} 
 Radial number densities $\rho (r)$ as a function of the 
 distance $r$ from the center of mass (CM) of the dendrimer for 
 $\lambda _{\rm B}$ equal to 0.0; 0.2\,$l_0$; 0.8\,$l_0$; and 
 8.0\,$l_0$. }
\end{figure*}

\begin{figure*}[tb] 
\includegraphics[width=10.3cm,clip=true]{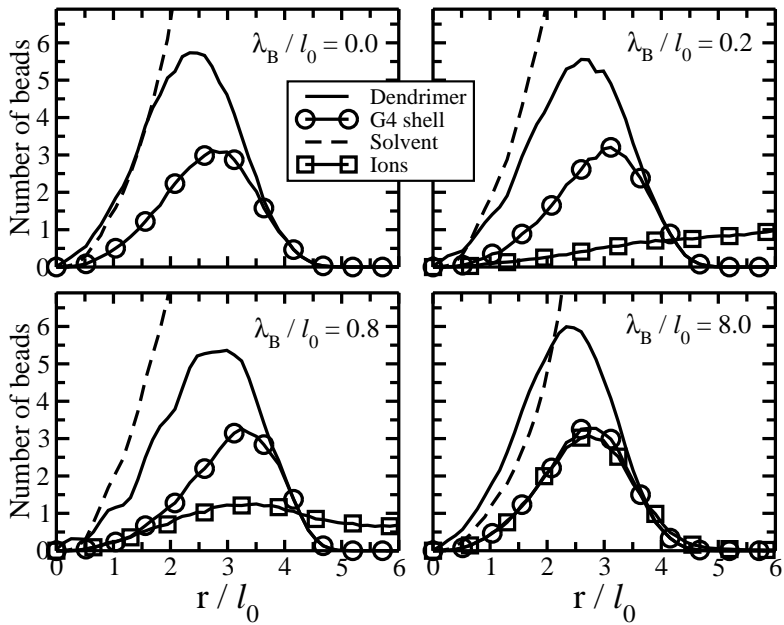}
\caption{\label{BeadNUM} 
 The number of beads (corresponding to the number densities 
 in Fig.~\ref{Density}) as a function of the distance $r$ from 
 the CM of the dendrimer.}
\end{figure*}

The $\lambda_{\rm B}$-dependence for the radius of gyration  
calculated for charged beads of the terminal shell 
($\langle R_{\rm g}^2 \rangle_{\rm term}$) follows 
closely that of the whole dendrimer 
($\langle R_{\rm g}^2 \rangle_{\rm dend}$), see Fig.~\ref{Rg}. 
Interestingly, the $\langle R_{\rm g}^2 \rangle_{\rm term}$ 
exceeds the radius of gyration of the dendrimer at all Bjerrum 
lengths. For neutral dendrimers this finding is in agreement 
with experimental data\cite{Topp99} and with Brownian dynamics 
simulations.\cite{LyulinA00a} As demonstrated in 
Ref.~\onlinecite{LyulinA00a}, and as we proceed to show, 
the fact that $\langle R_{\rm g}^2 \rangle_{\rm term} > 
\langle R_{\rm g}^2 \rangle_{\rm dend}$ does {\it not} mean 
that beads of the terminal generation shell are localized
near the dendrimer's periphery.

To clarify the role of counterions, we calculated the 
mean-square radius of gyration 
$\langle R_{\rm g}^2 \rangle_{\rm ions}$ of the 
counterions, see Fig.~\ref{Ions} (top). That serves as a
measure for the size of the counterion cloud around the 
dendrimer. It turns out that the ion cloud becomes smaller 
as the Bjerrum length increases. This decrease, however, 
is found to be monotonic. Furthermore, in Fig.~\ref{Ions} 
(bottom) we estimate the fraction of counterions condensed 
onto the dendrimer. For doing that we used a simple criterion, 
namely we counted counterions which had dendrimer beads 
in their first coordination shells. As Fig.~\ref{Ions} shows, 
the number of condensed ions increases continuously with 
the strength of electrostatic interactions in the system.

The above suggests that there is a subtle interplay between 
two opposite effects upon increasing $\lambda_{\rm B}$: 
({\it i}) A growth in the strength  of the {\it repulsive} 
electrostatic interactions between the charged terminal beads 
and ({\it ii}) a strengthening of condensation of counterions 
due to the {\it attractive} interactions between counterions 
and the oppositely charged dendrimer beads. The former leads 
to a swelling of the charged dendrimer and the latter 
screens positive charges of the terminal dendrimer generation 
giving rise to shrinking of the dendrimer. The maximum of the 
dendrimer's size is found to be at 
$\lambda_{\rm B}\simeq 0.8\,l_0$ (i.e. 
at $\lambda_{\rm B}\simeq \sigma$), see Fig.~\ref{Rg}.
This corresponds to swelling by $\sim$\,12.5\,\% with respect 
to the neutral case.

Condensation of counterions onto the dendrimer should be 
accompanied by a decrease in its hydration, which is 
characterized by the number of solvent molecules in the ion's 
first coordination shell. Indeed, that can be seen by computing 
the average coordination numbers of counterions with dendrimer's 
beads and with solvent molecules, see Fig.~\ref{CN}. The 
increase of the strength of electrostatic interactions in the 
system provokes a condensation of counterions and, correspondingly, 
leads to a loss of solvent molecules from the ions' first 
coordination (hydration) shell (Fig.~\ref{CN}).

\subsection{\label{density}Number density profiles\protect}

To analyze the locations of the dendrimer's beads, solvent 
molecules, and counterions in more detail, we calculated 
componentwise number density profiles $\rho (r)$ as a function 
of the radial distance $r$ from the center of mass (CM) of 
the dendrimer, see Fig.~\ref{Density} (Fig.~\ref{BeadNUM} 
shows the corresponding radial bead distributions). For 
a neutral dendrimer the density profiles are in agreement 
with earlier molecular dynamics studies of dendrimers in 
an explicit solvent.\cite{Karat01} The overall number density 
of a dendrimer develops a plateau at $r > 0.5\,l_0$ from its 
CM. The beads of the terminal shell are broadly distributed, 
implying a considerable degree of back-folding. The solvent 
molecules penetrate  deeply into the dendrimer and their 
density profile has a plateau  at $r > 0.5\,l_0$, 
see Fig.~\ref{Density}.

At small Bjerrum lengths ($\lambda_{\rm B}\leq 0.8\,l_0$) 
the charged dendrimer swells and the density profile plateau 
broadens considerably with a simultaneous drop in the plateau's 
magnitude, see, e.g., Fig.~\ref{Density} for 
$\lambda_{\rm B} = 0.8\,l_0$. The charged beads of the terminal 
shell are re-distributed towards the dendrimer's periphery 
making the interior of the dendrimer significantly more 
accessible for solvent and counterions.

Further increase in Bjerrum length leads to a shrinkage 
since counterion condensation starts to dominate. Therefore, 
the dendrimer's density profile changes towards a similar 
profile as observed at small $\lambda_{\rm B}$ with one 
important exception: the amount of condensed ions increases 
considerably, and their density profiles overlap more and 
more with the density profiles of the charged terminal beads 
of the dendrimer (up to almost complete coincidence of the 
profiles in the case of extremely large Bjerrum lengths, 
see Fig.~\ref{Density} for $\lambda_{\rm B}=8.0\,l_0$). 
This shrinkage is accompanied by a reduction of voids in the 
dendrimer's interior leading to squeezing of solvent molecules 
out of the dendrimer, see Figs.~\ref{Density} and \ref{BeadNUM}.

\subsection{\label{charge}Charge density profiles\protect}

\begin{figure}[tb] 
\includegraphics[width=8cm,clip=true]{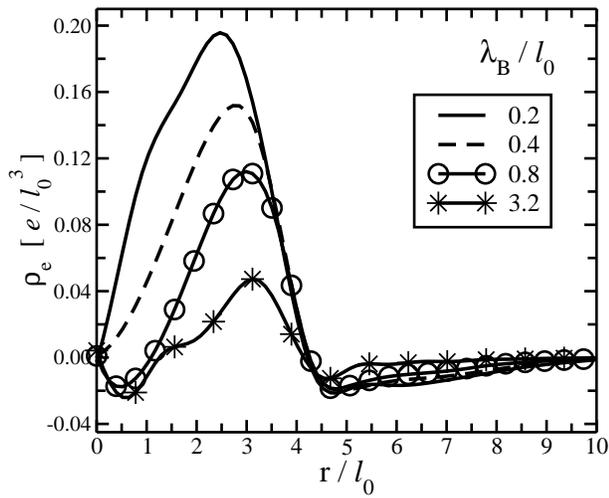}
\caption{\label{Charge} 
 Radial charge densities $\rho_{e}(r)$ as a function
 of the radial distance $r$ from the CM of a dendrimer for 
 $\lambda _{\rm B}$ equal to 0.2\,$l_0$ (solid line), 
 0.4\,$l_0$ (dashed line), 0.8\,$l_0$ (solid line with circles), 
 and 3.2\,$l_0$  (solid line with stars). To reduce the noise 
 in the data, the charge densities shown here were first
 fitted to splines.\cite{Thi98}}
\end{figure}

Next, we take a look at the screening of the dendrimer's 
charges by counterions. In Fig.~\ref{Charge} we plot the 
charge density profiles $\rho_{e}(r)$ for the whole system 
as a function of the radial distance $r$ from the CM of the 
dendrimer. We recall that since our solvent is non-polar, 
the charge densities $\rho_{e}(r)$ are defined only by the 
interplay between the positively charged terminal beads and 
the negatively charged counterions. Figure~\ref{Charge} 
clearly demonstrates screening: the height of the positive 
$\rho_{e}(r)$-peaks drops drastically with increasing 
Bjerrum length as compared to the $\rho_{e}(r)$-peak for 
$\lambda_{\rm B}=0.2\,l_0$ when the counterion 
condensation has just started.

While the main peaks of all $\rho_{e}(r)$-curves are located 
in the domain of the positive charges, starting with 
$\lambda_{\rm B}=0.6\,l_0$, one observes small minima close 
to the CM of the dendrimer where $\rho_{e}(r)$ becomes negative, 
see Fig.~\ref{Charge}. We address these minima to the fact that 
some of condensed counterions can be localized deep in the 
dendrimer's interior rather than beside charged terminal beads 
giving rise to an excess of negative charges close to the 
dendrimer's center of mass, see Figs.~\ref{Density} and 
\ref{BeadNUM}. In turn, on the right of $\rho_{e}(r)$-peaks 
we see another region where the overall charge of the system 
is also negative. This region of negative charge, being 
associated with ``unbound'' counterions in bulk solvent, 
becomes smaller with $\lambda_{\rm B}$ and eventually 
disappears at large Bjerrum lengths.

\section{\label{dynamics}Dynamic properties\protect}

The molecular dynamics simulation technique provides an 
access to the dynamic properties, and next we discuss the 
dynamics of charged dendrimers in a salt-free solution.

\subsection{\label{global}Global reorientational dynamics\protect}

First, we consider the relaxation of vectors pointing from 
the center of mass (CM) of the dendrimer to the beads of 
a given generation shell.\cite{Cai97,Karat01,LyulinS04a} 
If the terminal (G4) generation shell is considered, the 
corresponding autocorrelation function (ACF) of ``CM-G4'' 
vector can be employed to analyze the global reorientational 
behavior of the whole dendrimer.\cite{Karat01} It is, 
however, necessary to bear in mind that there is no reason 
for the beads of the last generation to be located at the 
dendrimer's periphery.

In general, relaxation of the ``CM-G4'' vector is determined
by the rotation of the whole dendrimer together with fluctuations 
in the vector's length. To characterize the dendrimer's rotation, 
we calculated the ACF of the {\it unit} vector directed along 
the ``CM-G4'' vector. It turns out that the corresponding 
relaxation times $\tau_{\rm rot}$ are very close to the 
relaxation time of the "CM-G4" vector ($\tau_{\rm shell}$). 
Thus, the relaxation of the ``CM-G4'' vector is governed almost 
exclusively by the rotation of the dendrimer in agreement with 
the results from Brownian dynamics simulations of dendrimers 
with rigid bonds.\cite{LyulinS04a}

\begin{figure}[tb] 
\includegraphics[width=8cm,clip=true]{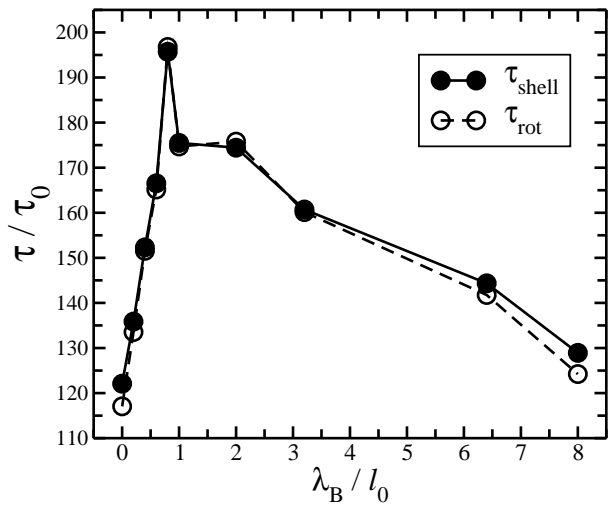}
\caption{\label{TauREO} 
 Characteristic relaxation times $\tau_{\rm shell}$ and 
 $\tau_{\rm rot}$ related to the dendrimer's reorientational 
 dynamics as extracted from the corresponding autocorrelation 
 functions of vectors pointing from the CM of the dendrimer 
 to the beads of its terminal G4 generation shell, see text 
 for details. Error bars are of the order of the symbol size.
}
\end{figure}

Figure~\ref{TauREO} shows that the global reorientational 
motion of the charged dendrimer depends non-monotonically 
on the Bjerrum length. Such a behavior resembles the  
$\lambda_{\rm B}$-dependence of $\langle R_{\rm g}^2 \rangle$ 
as was seen earlier in Fig.~\ref{Rg}. This is expected since 
it is reasonable to assume that the global reorientational 
dynamics is controlled by the dendrimer's size: larger size 
requires more time for reorientation.

As Fig.~\ref{TauREO} shows, there is a sharp maximum
at $\lambda_{\rm B}\simeq 0.8\,l_0$. It is also interesting 
that although at large Bjerrum lengths the average size of 
the charged dendrimer is found to be smaller than the size 
of a neutral dendrimer (Fig.~\ref{Rg}), that is not reflected 
in the characteristic relaxation times shown in Fig.~\ref{TauREO}.
The former can be explained by the inhomogeneous distribution of
counterions inside the dendrimer, giving rise to the additional electrostatic attraction
between different parts of the dendrimer interior. On the other hand, the 
global reorientation dynamics is not only governed by the 
dendrimer's size but also by its mass. At 
$\lambda_{\rm B}= 8.0\,l_0$ almost all counterions are bound 
to the dendrimer (Fig.~\ref{Ions}), thus increasing the 
dendrimer's effective mass by a factor of 1.5 in comparison 
to the neutral one. This, in turn, slows down the global 
reorientational motion of a dendrimer.

\begin{figure}[tb] 
\includegraphics[width=8cm,clip=true]{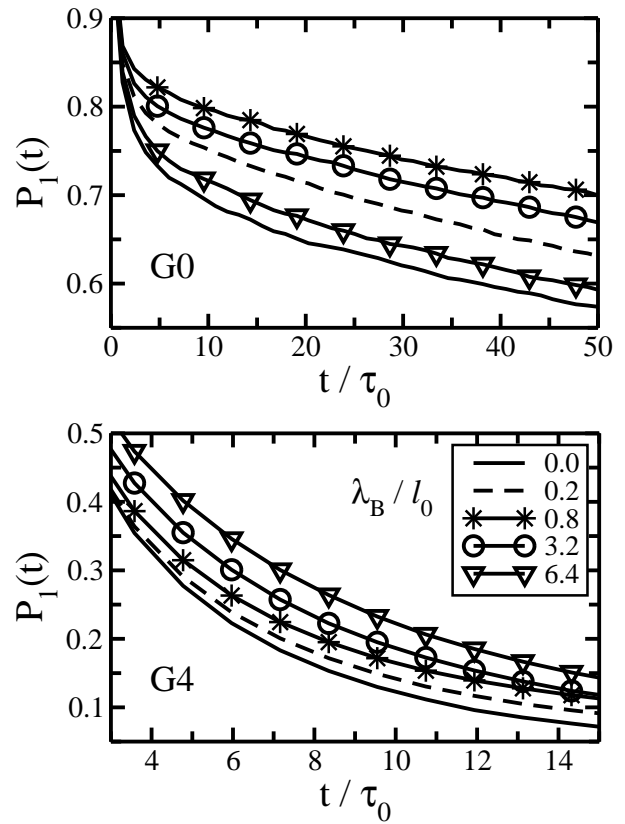}
\caption{\label{P1} 
 Time evolution of the normalized autocorrelation function 
 $P_1 (t) = \langle {\bf b}_i(0){\bf b}_i(t) \rangle$ of bond 
 vectors ${\bf b}_i(t)$ of the innermost ({\it top}) and the 
 outermost ({\it bottom}) shells. The legend shows the 
 different Bjerrum lengths used.
}
\end{figure}

\subsection{\label{local}Local dynamics of bonds and dendrimer's core\protect}

As seen in Fig.~\ref{TauREO}, the global reorientational 
dynamics is determined by rather long characteristic times 
up to $\sim$\,200\,$\tau_0$. The local orientational motions 
are characterized by much shorter times. As an example, we 
consider here the re-orientation dynamics of the bond vectors 
${\bf b}_i(t)$. In Fig.~\ref{P1} we plot the time evolution 
of the corresponding autocorrelation function 
$P_1 (t) = \langle {\bf b}_i(0){\bf b}_i(t) \rangle$ for 
the innermost (G0) and the outermost (G4) shells. We note 
that the relaxation of $P_1 (t)$ cannot be characterized by 
a single relaxation time. Therefore, we focus here on the 
time behavior of $P_1 (t)$ rather than on characteristic times.

The reorientation of a bond slows down with the size of the 
dendrimer, see Fig.~\ref{P1} (top). Such a behavior is found 
to hold for all inner generation shells from G1 to G3 (data 
not shown). Remarkably, the above non-monotonic (notice the 
order of the curves) dependence on $\lambda_{\rm B}$ in 
Fig.~\ref{P1} (top) breaks 
down for $P_1 (t)$ of the terminal (outermost) generation 
shell as it is seen in Fig.~\ref{P1} (bottom). This feature 
can be linked to condensation of counterions which plays 
a dominant role at $\lambda_{\rm B} > 0.8\,l_0$ and, therefore, 
can hinder the reorientation of bonds of the terminal G4 shell.

\begin{figure}[tb] 
\includegraphics[width=8cm,clip=true]{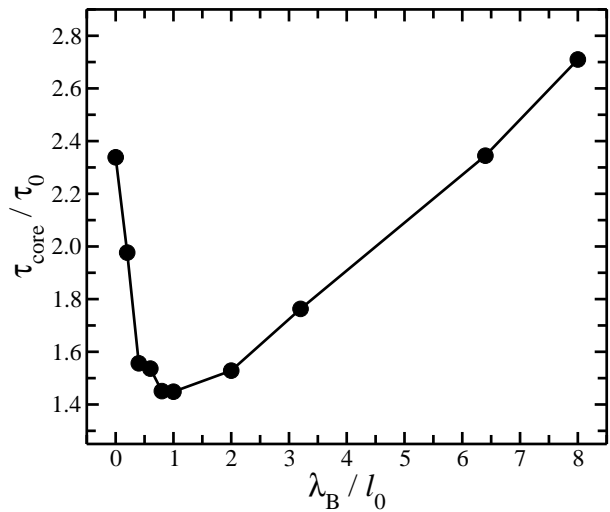}
\caption{\label{CM} 
 The characteristic relaxation time $\tau_{\rm core}$ for 
 a vector pointing from the CM of a dendrimer to its core 
 as a function of the Bjerrum length $\lambda_{\rm B}$. 
 Error bars are of the order of symbol size.}
\end{figure}

The local re-orientation dynamics of dendrimer bonds
can also be characterized by the autocorrelation function 
$P_2 (t) = (3/2)\big(\langle {\bf b}_i(0){\bf b}_i(t) 
\rangle ^2 -1/3 \big)$. It turns out that the simple 
relation $P_2 (t)=P_1^3 (t)$ holds for all bonds  
in agreement with the previous simulation studies.\cite{LyulinS04a}

Last, we focus on the relaxation of the vector pointing from
the dendrimer's center of mass to its core (``CM-core'' vector). 
\cite{LyulinS04a} The core and the CM are typically found to 
be very close to each other. However, they do not exactly 
coincide and the relaxation of the corresponding ``CM-core'' 
vector is characterized by small (but finite) relaxation times.

In Fig.~\ref{CM} we plot the corresponding characteristic 
relaxation times $\tau_{\rm core}$. As argued above, the 
relaxation times turn out to be very small, just a few $\tau_{\rm 0}$. 
The $\lambda_{\rm B}$-dependence of the $\tau_{\rm core}$ 
resembles the behavior of the $\langle R_{\rm g}^2 \rangle$-curve
in Fig.~\ref{Rg}, i.e., increase in the dendrimer size speeds 
up the relaxation of the ``CM-core'' vector and {\it vice versa}. 
This effect can be explained in terms of the mobility of 
a dendrimer core: larger dendrimers are characterized by 
smaller densities of their interior (see Fig.~\ref{Density}) 
and, therefore, by more mobile cores which relax faster.

\section{\label{end}Discussion and Conclusions\protect}

We have performed a systematic molecular dynamics study 
of generic cationic dendrimers solvated in {\it explicit} 
solvent in the presence of {\it explicit} counterions. 
The main goal was to understand to what degree the structural 
and dynamic properties of charged dendrimers are sensitive 
to the strength of electrostatic interactions, and what is 
the role of counterions.

We found an intriguing interplay between the {\it repulsive} 
interactions of the dendrimer's charged terminal beads and 
their {\it attractive} interactions with oppositely charged 
counterions. Depending on the type of dominant interactions, 
one observes swelling or shrinking of the molecule as the 
strength of the electrostatic interactions is increased. In 
other words, the size of a charged dendrimer depends 
{\it non-monotonically} on the Bjerrum length $\lambda_{\rm B}$ 
as shown in Fig.~\ref{Rg}. Most of the dendrimer's structural 
and dynamic characteristics considered in this study were 
found to follow this pattern.

For small Bjerrum lengths a charged dendrimer swells with  
$\lambda_{\rm B}$ and its size reaches maximum at 
$\lambda_{\rm B}=0.8\,l_0$. At larger Bjerrum lengths the 
condensation of counterions dominates and the dendrimer 
shrinks. The counterion condensation is often discussed 
in the framework of the Manning theory,\cite{Manning78} 
which predicts the condensation to occur when 
$\lambda_{\rm B}/a \geq 1$ where $a$ is the distance 
between charges along the polymer backbone. However, the 
Manning's criterion is not applicable in our case. The 
Manning's theory is formulated for linear polyelectrolytes 
and not for spherical-like branched objects such as dendrimers.
Second, the charged dendrimer considered in our study has 
only its terminal beads charged, making the meaning of 
the parameter $a$ uncertain. The counterion condensation, 
the effects of valence, charging, and the size and shape 
of the dendrimer could more appropriately be studied using 
the method of Patra~et~al.\cite{Patra:03rw} who studied 
the charging of spherical three dimensional objects under 
different conditions. A detailed study of the above factors 
is, however, beyond the current work and will be addressed 
in a future publication.

The Bjerrum length $\lambda_{\rm B} = \sigma$ corresponds to 
the position of the $\langle R_{\rm g}^2 \rangle$-peak, see 
Fig.~\ref{Rg}, and, therefore, separates swelling and shrinking 
regimes. Remarkably, on the left-hand side of the peak, the 
separation between charges {\it always} exceeds the Bjerrum 
length because the beads cannot get closer than $\sigma$ and 
$\lambda_{\rm B} < \sigma$. On the other hand, the counterion 
condensation is rather weak in this $\lambda_{\rm B}$-domain. 
Therefore, one can expect the Debye--H{\"u}ckel approximation 
to be valid at around $\lambda_{\rm B}\leq\sigma$ and to break 
down at larger Bjerrum lengths.

It is instructive to relate the dendrimer bond length $l_0$ 
(and, correspondingly, the average dendrimer size $R_{\rm g}$ 
and the Bjerrum length $\lambda_{\rm B}$) to the characteristic 
length scales typical for real dendrimers in aqueous solutions. 
The value of 17.1\,\AA\, is available from small angle X-ray 
scattering experiments\cite{Prosa97} for the average radius 
of gyration $R_{\rm g}$ of a PAMAM dendrimer of the fourth 
generation. A very recent MD study\cite{Maiti05} reported 
$R_{\rm g} = 16.78 \pm 0.15$\,\AA\, for a non-protonated PAMAM 
dendrimer in explicit water. In another study, Lee et al. 
\cite{Lee02} reported $R_{\rm g} = 14.8\pm 0.1$\,\AA. That 
was obtained from MD studies of PAMAM dendrimers without solvent 
($R_{\rm g} = 14.50 \pm 0.28$\,\AA\, in Ref.~\onlinecite{Maiti05} 
was obtained under the same conditions). Assuming that dendrimer 
swells on about 15\,\% in water,\cite{Maiti05} the results of 
Lee et al.\cite{Lee02} seem to be in accord with other studies. 
We took $R_{\rm g}\simeq 17$\,\AA\, for our parameterization. 
For a neutral dendrimer we have $R_{\rm g}\simeq 2.54~l_0$, 
see Fig.~\ref{Rg}, such that $l_0\simeq 6.7$\,\AA.

Thus, the critical Bjerrum length corresponding to the 
dendrimer's maximal size can be estimated to be 5.4\,\AA. 
In turn, the  Bjerrum length in water at 300 K is about 
7.1\,\AA\, (or $\sim 1.3\,l_0$), i.e., it exceeds the critical 
value. In practice, since the $\langle R_{\rm g}^2 \rangle$-peak 
is not very sharp, the value of $\langle R_{\rm g}^2 \rangle$ 
for a dendrimer in {\it water}, being located next to the 
peak, is rather close to the maximum of 
$\langle R_{\rm g}^2 \rangle$, see Fig.~\ref{Rg}. Taking also 
into account the uncertainty in defining the value of $l_0$, 
one can conclude that counterion condensation  will not affect 
significantly PAMAM dendrimers in water under physiological 
conditions.

To summarize, we demonstrated that including explicit 
counterions can have a dramatic effect on the structure and 
dynamics of charged dendrimers, and under certain conditions 
they cannot be treated implicitly. Therefore, simplified 
approaches, such as the well-known Debye--H{\"u}ckel 
approximation, have to be applied with great care. Given 
that the most exciting applications of protonated dendrimers are 
related to drug and gene delivery, it would be interesting 
to clarify the role of counterions in complexation of charged 
dendrimers with linear polyelectrolytes bearing opposite charges. 
\cite{Welch00,LyulinS05} This problem is addressed in our 
ongoing studies.

\begin{acknowledgments}

Fruitful discussions with Dr. Alexey V. Lyulin are gratefully 
acknowledged. This work has been supported by the Academy of 
Finland through its Center of Excellence Program (A.\,A.\,G., 
I.\,V.) and through Grants 202598 (A.\,A.\,G.), 80246 (I.\,V.), 
00119 (M.\,K.), 54113 (M.\,K.), and 211579 (S.\,V.\,L.) and 
by the Russian Foundation of Basic Research through Grants 
05-03-32332  (A.\,A.\,G.) and 05-03-32450-a (S.\,V.\,L.). 
S.\,V.\,L. acknowledges the support of the Government of 
St.\,Petersburg, Russia (Grant PD 05-1.3-101) and M.\,K. 
the support from Emil Aaltonen Foundation (Finland). The 
simulations were performed on the HorseShoe (DCSC) supercluster 
at the University of Southern Denmark.
\end{acknowledgments}


\end{document}